# Strain-tunable direct band gap of ZnO monolayer in graphene-like honeycomb structure

Harihar Behera[1] and Gautam Mukhopadhyay[2]

Department of Physics, Indian Institute of Technology Bombay, Powai, Mumbai-400076, India

**ABSTRACT**

Using full-potential density functional calculations within local density approximation (LDA), we found mechanically tunable band-gap in ZnO monolayer (ML-ZnO) in graphene-like honeycomb structure, by simulated application of in-plane homogeneous biaxial strain. Unstrained ML-ZnO was found to have a direct band gap of energy 1.68 eV within LDA; the actual band gap would be more, since LDA is known to underestimate the gap. Within our simulated strain limit of about plus or minus 10%, the band gap remains direct and shows a strong non-linear variation with strain. The results may find applications in future nano-electromechanical systems (NEMS) and nano-optomechanial systems (NOMS).

---

[1] E-mail: harihar@phy.iitb.ac.in; harihar@iopb.res.in

[2] Corresponding author; E-mail: gmukh@phy.iitb.ac.in

Nanomaterials whose electronic properties can be controlled by mechanical strain are greatly desired for applications in nano-electromechanical systems (NEMS). If such materials happen to have a direct band gap, then those can find applications in nano-optomechanical systems (NOMS). If mechanical strain strongly affects the electronic properties of a material, then that material can be used as a strain sensor. Bulk ZnO, in its most stable wurtzite structure at ambient pressure, has many useful properties[1,2] such as direct and wide band gap ($E_g$ = 3.37 eV at room temperature and 3.44 eV at low temperatures), large exciton binding energy (60 meV), strong piezoelectric and pyroelectric properties, strong luminescence in the green-white region, large non-linear optical behavior, high thermal conductivity, radiation hardness, biocompatibility, and so on. These properties make ZnO a great candidate for a variety of applications in optics, electronics and photonics, sensors, transducers and actuators. Several ZnO nanostructures in the form of very thin nanosheets, nanowires, nanotubes and nanobelts have been synthesized and characterized using different preparation methods.[3,4] In 2006, using density functional theory (DFT) calculations, Freeman et al.[5] predicted that when the layer number of ZnO films with (0001) orientation is small, the wurtzite structures are less stable than a phase based on two dimensional (2D) ZnO sheets with a layer ordering akin to that of hexagonal BN. In 2007, Tusche et al.[6] reported the observation of 2 monolayer (ML) thick ZnO(0001) films grown on Ag(111) by using surface x-ray diffraction and scanning tunneling microscopy. Very recently graphene-like honeycomb structures of ZnO have been prepared[7] on Pd(111) substrate. Theoretical studies[5,8-10] have now established the stability of the monolayer and few-layers (FL) of ZnO in planar honeycomb structures. This stability is attributed to the strong in-plane sp$^2$ hybridized bonds between the Zn and O atoms.

Recent theoretical studies reported (i) the elastic, piezoelectric, electronic, and optical properties[10] of ML-ZnO, (ii) room-temperature half-metallic ferromagnetism in the half-fluorinated[11] ML-ZnO, (iii) fluorination[12] induced half metallicity in few ZnO layers, (iv) fully-fluorinated and semi-fluorinated[13] ZnO sheets and (v) strain-induced semiconducting-metallic transition[14] for ZnO zigzag nanoribbons. Here, we report our DFT based investigation of the effect of homogeneous biaxial strain on the electronic properties of ML-ZnO.

The calculations have been performed by using the DFT based full-potential (linearized) augmented plane wave plus local orbital (FP-(L)APW+lo) method,[15] which is a descendant of FP-LAPW method.[16] We use the elk-code[17] and the Perdew-Zunger variant of LDA,[18] the accuracy of which have been successfully tested in our previous works on graphene and silicene[19] (silicon analog of graphene)[20,21] and germanene[22] (germanium analog of graphene).[20,21] The plane wave cut-off of $|G+k|_{max}= 9.0/R_{mt}$ (a.u.$^{-1}$) ($R_{mt}$ = the smallest muffin-tin radius) was chosen for plane wave expansion in the interstitial region. The Monkhorst-Pack[23] k-point grid of 30×30×1 was used for all calculations. The total energy was converged within 2μeV/atom. We simulate the 2D-hexagonal structure of ZnO-ML as a 3D-hexagonal supercell with a large value of $c$-parameter (= |**c**| = 40 a.u.). The application of homogeneous in-plane biaxial strain δ ≈ ± 10% was simulated by varying the in-plane lattice parameter $a$ (=|**a**| = |**b**|); δ = $(a − a_0)/a_0$, where $a_0$ is the ground state in-plane lattice constant. Fig. 1 depicts the top-down view of a ZnO monolayer in graphene-like planar honeycomb structure.

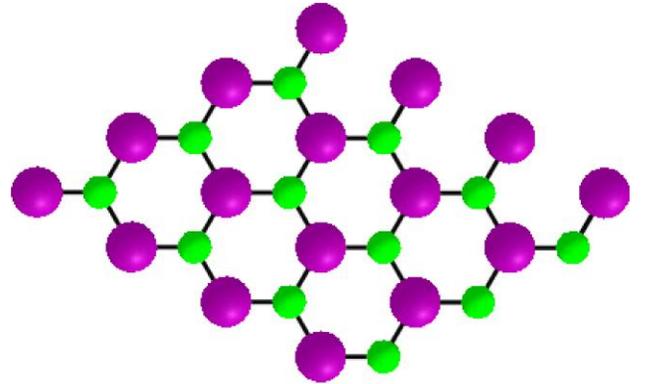

FIG. 1. Top-down view of a ZnO monolayer in graphene-like honeycomb structure in ball-stick model. The large (pink) ball represents Zn atom and the small (green) ball represents the O atom.

Our calculated LDA value of $a_0$ = 3.20 Å corresponding to the Zn-O bond length $d_{Zn-O}$ =

$a_0/\sqrt{3}$ = 1.848 Å, which is in agreement with reported the experimental[6] value of $d_{Zn-O}$ = 1.92 Å, and theoretical values of $d_{Zn-O}$ = 1.86 Å by Tu and Hu[8], 1.895 Å by Topsakal et al.[9], 1.853 Å by Tu[10], 1.85 Å by Wang.[12] Our 3.75% underestimation of $d_{Zn-O}$ value with respect to the experimental value is due to the well known problem of underestimation of the lattice constant within LDA. In our previous studies[18,19] of graphene, silicene and germanene, we have demonstrated that the value of $c$-parameter chosen in the construction of supercell for simulation of 2D hexagonal structures also affects the value of in-plane lattice constant: larger value of $c$-parameter yields a slight smaller value of $a$. Since we use a different method and a different value of $c$-parameter the disagreement of our result on $a_0$ with other theoretical result,[9] is acceptable.

The electronic band structure (Bands) and total density of states (TDOS) plots of unstrained ML-ZnO are depicted in Fig. 2. As seen in Fig. 2, ML-ZnO is a direct band gap ($E_g$ = 1.68 eV, LDA value) semiconductor with both valence band maximum (VBM) and conduction band minimum located at the $\Gamma$ point of the hexagonal Brillouin Zone. However, the actual band gap is expected to be larger as LDA is known to underestimate the gap. Our calculated LDA band gap of 1.68 eV is in agreement with previous calculations of 1.68 eV by Topsakal et al.,[9] 1.762 eV by Tu,[10] 1.66 eV by Wang et al.[13] Using the GW approximations, recently Tu[10] estimated the direct band gap of ML-ZnO at 3.576 eV, which showed ML-ZnO as a wide band gap semiconductor.

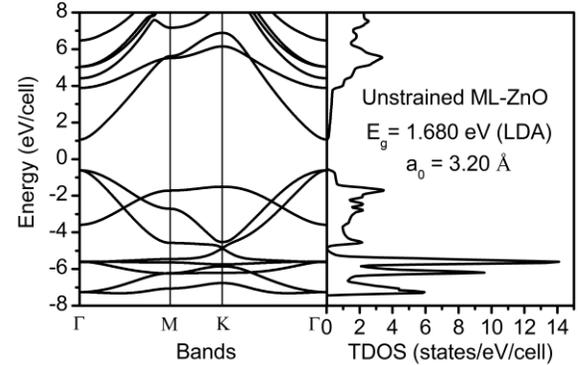

FIG. 2. Bands and TDOS of unstrained ML-ZnO within LDA. Fermi energy $E_F$ = 0 eV.

Our calculated results on homogeneous biaxial strain-induced modifications of the band gap of ML-ZnO are depicted in Fig. 3, which shows a strong non-linear variation of band gap with biaxial strain. Within our simulated strain limits as shown in Fig. 3, the band gap remains direct. This type of strain-engineered band gap of ML-ZnO, if verified by experiments, may be useful in fabrication of NEMS and NOMS based on a graphene-analogue of ZnO.

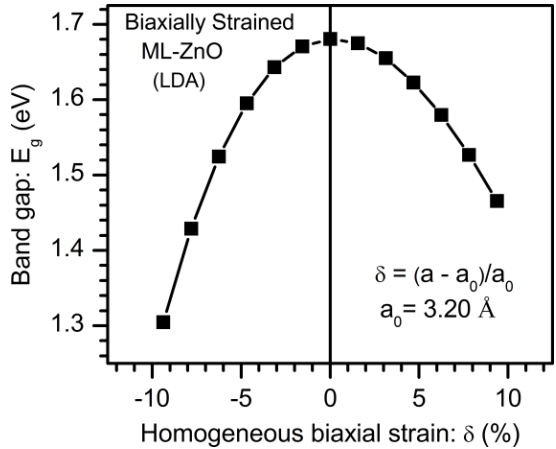

FIG. 3. Variation of the direct band gap of ML-ZnO with homogeneous biaxial strain $\delta$.

In summary, we found a mechanically tunable band-gap in ZnO monolayer (ML-ZnO) in graphene-like honeycomb structure, by simulated application of in-plane homogeneous biaxial strain. Owing to its direct band gap with strong non-linear variation with biaxial strain, ML-ZnO may have potential applications in mechatronics and optical nanodevices such as NEMS and NOMS.